\documentclass[aps,prx,preprint,showpacs]{revtex4-1}
\usepackage[T1]{fontenc}
\usepackage[latin9]{inputenc}
\usepackage{hyperref}
\usepackage{amsmath}
\usepackage{amssymb}
\usepackage{graphicx}
\usepackage{subscript}
\usepackage{color}
\makeatletter
\begin{document}
%%%%%%%%%%%%%%%%%%%%%%%%%%%%%% LyX specific LaTeX commands.

\title{Interaction Effects on the Magneto-optical Response of Magnetoplasmonic
Dimers}

\author{N. de Sousa}
\affiliation{Departamento de F\'\i sica de la Materia Condensada, Universidad Aut\'onoma
de Madrid, E-28049 Madrid, Spain}
\author{L. S. Froufe-P\'erez}
\affiliation{Instituto de Estructura de la Materia, CSIC, Serrano 121, E-28006
Madrid, Spain}
\author{G. Armelles}
\author{A. Cebollada}
\author{M. U. Gonz\'alez}
\author{F. Garc\'\i a}
\author{D. Meneses-Rodr\'\i guez}
\author{A. Garc\'\i a-Mart\'\i n}
\email{a.garcia.martin@csic.es}
\affiliation{IMM-Instituto de Microelectr\'onica de Madrid (CNM-CSIC), Issac Newton
8, PTM, E-28760 Tres Cantos, Madrid, Spain}

%%%%%%%%%%%%%%%%%%%%%%%%%%%%%% User specified LaTeX commands.

\makeatother

\begin{abstract}
The effect that dipole-dipole interactions have on the magneto-optical
(MO) properties of magnetoplasmonic dimers is theoretically
studied. The specific plasmonic versus magnetoplasmonic nature of
the dimer\textquoteright{}s metallic components and their specific
location within the dimer plays a crucial role on the determination
of these properties. We find that it is possible to generate an induced
MO activity in a purely plasmonic component, even larger than that
of the MO one, therefore dominating the overall MO spectral dependence
of the system. Adequate stacking of these components may allow obtaining,
for specific spectral regions, larger MO activities in systems with
reduced amount of MO metal and therefore with lower optical losses.
Theoretical results are contrasted and confirmed with experiments for selected structures.
\end{abstract}


\maketitle

%\section{Introduction}

Smart nanoscale systems are able to interact with light in an intricate
fashion, \cite{aeschlimann_adaptive_2007} which is strongly dependent
on the internal electromagnetic interaction between the constituent
elements of the system. Plasmonic structures composed by a number
of individual elements, for example, give rise to Fano resonance effects
that induce Electromagnetically Induced Transparency (EIT). \cite{hentschel2010transition,lassiter2010fano,su2003interparticle,rechberger2003optical,dmitriev2007gold,brown2010heterodimers,wadell2012optical}
Similar phenomena have also been found in magnetoplasmonic nanosystems,\cite{armelles2013magnetoplasmonics}
those sharing magnetic and plasmonic functionalities and that therefore
allow a further degree of freedom, namely the external control of
the system response. \cite{temnov_active_2010,belotelov_enhanced_2011,bonanni_designer_2011,banthi_high_2012,chin_nonreciprocal_2013}
By an adequate design of their internal structure, it is possible
to obtain configurations which provide enhanced MO activity upon plasmon
resonance excitation, \cite{gonzalez2008plasmonic,jain2009surface,wang2011plasmonics,marinchio2012light}
which allow probing the EM field distribution inside a metallic nanoelement,
\cite{meneses2011probing} or which yield high MO activity and low
optical losses with MO figures of merit comparable with those of garnet
structures.\cite{banthi_high_2012} Furthermore, in dimers where
one of the elements is purely plasmonic and the other is of magnetoplasmonic
nature, interaction effects cause the magnetoplasmonic component to
induce MO activity in the plasmonic one (which intrinsically lacks
MO activity).\cite{armelles2013mimicking} For specific inter-element
distances, which determines the interaction between them, this brings
as a consequence the equivalent of the EIT in the MO spectrum of the
system, \textit{i.e.} a cancelation of the MO activity in a narrow
spectral range due to the competition between the intrinsic MO contribution
of the magnetoplasmonic component and the induced MO contribution
of the plasmonic one.\cite{armelles2013mimicking} As this effect
exhibits a narrow spectral feature in the MO response, it may find
applications in sensing and telecommunication areas, and a complete
understanding will help to the development of novel sensing and biosensing
architectures as well as MO devices.

In this context, these induced MO activity effects and its influence
on the overall MO activity of the system for specific ranges of interaction
lead to consider additional issues where the electromagnetic interaction
between these elements is relevant but remains unaddressed. For example:
Is it possible to devise a configuration for which the MO activity
induced in the non MO-active element is even larger than that of the
MO active one? Even more, does the MO response depend in a continuous,
gradual fashion with the amount of MO active component? Moreover,
in systems where both components are MO active, does the MO response
behaves simply as the sum of those of the two components?

With this in mind, the goal of this work is to consider theoretically
and experimentally these issues by presenting a detailed study of
the interaction effects in a model system formed by two coupled nanodisks
separated by a dielectric in a nanopillar geometry when the plasmonic
or magnetoplasmonic nature of the nanodisk components is changed.
Namely, we present results for three different geometries: first assuming
that the bottom disk is magnetoplasmonic and the top one is plasmonic;
second the inverse situation (i.e. top magnetoplasmonic, bottom plasmonic);
and finally the case in which both nanodisks are magnetoplasmonic
in nature. For the theoretical description we will follow two approaches,
an analytic one in which each disk is considered as a point dipole
(with the proper polarizability) and a numerical one based on FDTD
techniques in which the real internal structure of the disks is taken
into account. The first, simple approach allows distinguishing the
contribution of each of the elements separately, giving detailed information
about the underlying physics. The second, full numerical approach,
permits the validation of the obtained insights. These theoretical
results will be contrasted with experimental data of equivalent systems
obtained by hole mask colloidal lithography and evaporation.

%\section{Analytical approach: two interacting dipoles }

\begin{figure}
\includegraphics[width=1\columnwidth]{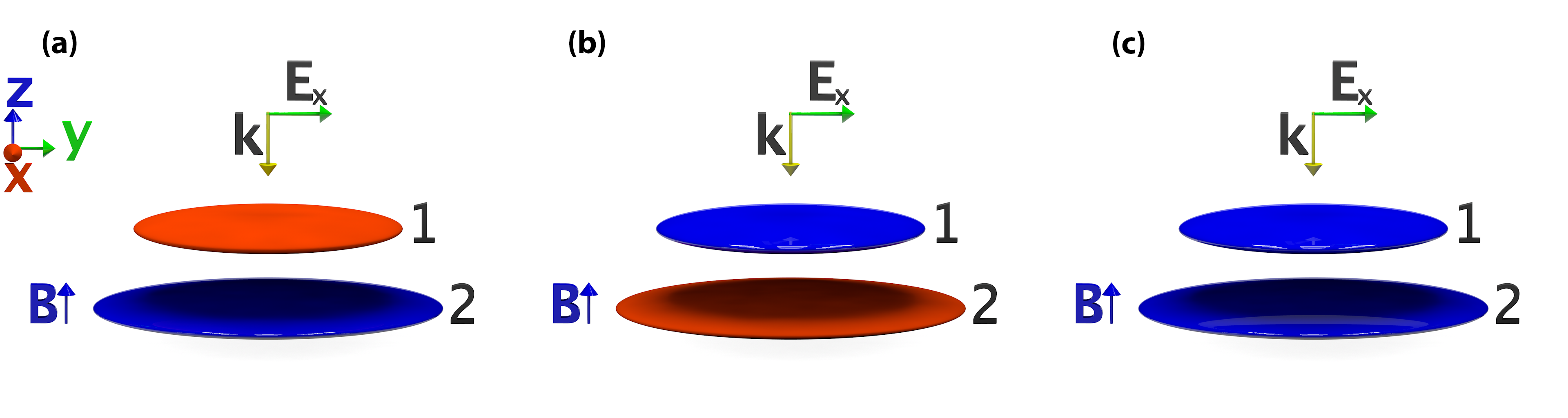}

\caption{Color online. Schematic representation of the studied configurations.\protect \\
 (a) The lower dipole is MP whereas the upper one is P, (b) the lower
dipole is P whereas the upper disk is MP, (c) both disks are MP.\label{schemetic_representation}}
\end{figure}
The geometry is similar to that previously considered in Refs. \cite{dmitriev2007gold,armelles2013mimicking,banthi_high_2012},
where two metallic disks (each one can be magneto-optically active)
are vertically aligned and separated by a dielectric spacer. We approximate
each disk by an oblate spheroid with an aspect ratio that corresponds
to dimensions of previously fabricated disks (see Fig. \ref{schemetic_representation}).
Since the actual fabricated structures have a truncated cone shape,
the aspect ratio of the bottom dipole must be larger than that of
the top one. For non-magneto-optical, plasmonic dipoles (P) we consider
a diagonal, isotropic, dispersive dielectric tensor (Au). For the
magnetoplasmonic (MP) dipole we consider an average medium that combines
the dielectric tensor of a noble metal and that of a ferromagnetic
one (Au and Co in this case) giving rise to a non-diagonal dielectric
tensor. The non-zero off-diagonal elements depend on the relative
orientation of the geometry, of the exciting radiation and of the
magnetic field. In our case the external magnetic field is aligned
perpendicular to the dipoles (i.e. aligned along the stacking direction,
see Fig. \ref{schemetic_representation}), which corresponds to
the so called polar Kerr configuration, and the dielectric tensor
of the MP dipole presents the form:

\noindent 
\begin{equation}
\boldsymbol{\varepsilon}=\begin{pmatrix}\varepsilon_{\text{d}} & \varepsilon_{\text{M}} & 0\\
-\varepsilon_{\text{M}} & \varepsilon_{\text{d}} & 0\\
0 & 0 & \varepsilon_{\text{d}}
\end{pmatrix}.\label{Tensor}
\end{equation}

\noindent Depending of the amount of Co within the MP disk, the elements
of the dielectric tensor read as: 

\noindent 
\begin{equation}
\boldsymbol{\varepsilon}_{\text{{d}}}=\left(1-\nu\right)\boldsymbol{\varepsilon}_{\text{d,Au}}+\nu\boldsymbol{\varepsilon}_{\text{d,Co}};\:\boldsymbol{\varepsilon}_{\text{{M}}}=\nu\boldsymbol{\varepsilon}_{\text{M,Co}}\label{Concent}
\end{equation}

\noindent where $\nu=\frac{V_{Co}}{V_{Co}+V{}_{Au}}$ is the Co relative
amount in each dipole.

Once the dielectric tensor is known, we can obtain the static polarizability
of a dipole, that, considered as an oblate particle in air, is given
by:\cite{bohren1998book}

\noindent 
\begin{equation}
\boldsymbol{\tilde{\alpha}}_{0}=4\pi a^{2}c\frac{\boldsymbol{\varepsilon}-\mathbb{I}}{3\mathbb{I}+3\mathbf{L}\left(\boldsymbol{\varepsilon}-\mathbb{I}\right)},
\end{equation}
where $a$ and $c$ are the in-plane and out of plane dimensions of
the oblate spheroids (see Fig. \ref{schemetic_representation}),
$\text{\ensuremath{\varepsilon}}$ is the dielectric tensor of the
material and $\mathbf{L}$ the geometrical tensor. To insure the optical
theorem is fulfilled, we apply the radiative correction to the static
polarizability:\cite{albadalejo2010radiative}

\noindent 
\begin{equation}
\boldsymbol{\tilde{\alpha}}=\frac{\boldsymbol{\tilde{\alpha}}_{0}}{\mathbb{I}-i\frac{k^{3}}{6\pi}\boldsymbol{\tilde{\alpha}}_{0}}.\label{Pola}
\end{equation}

Sometimes it is convenient to work with scaled magnitudes, so that
polarizablity $\tilde{\alpha}$, polarization $\tilde{p}$ and green
tensor $\tilde{G}$, become $\alpha_i=(k^3/4\pi)\tilde{\alpha_i}$,
$p_i=(k^3/4\varepsilon_0\pi)\tilde{p_i}$, $G=(4\pi/k)\tilde{G}$.
Once we have the polarizability for oblate particles, we are able
to describe each disk as a single dipole. Its properties (material,
shape and dimensions) are embedded in the polarizability. From coupled
dipole theory we know that the interaction between dipoles is mediated
by the Green tensor $G$. If an incident planar wave, with
wavenumber $k$ and with electric polarization in the plane of the
dipoles, is used to excite the system (see Fig. \ref{schemetic_representation})
the two dipoles can be described in the x-y plane as: 
\begin{eqnarray}
\mathbf{p}_{1} & = & \boldsymbol{\alpha}_{1}\left[\mathbf{E}_{0,1}+{G}\left(\mathbf{r}_{1},\mathbf{r}_{2}\right)\mathbf{p}_{2}\right]\nonumber \\
\mathbf{p}_{2} & = & \boldsymbol{\alpha_{2}}\left[\mathbf{E}_{0,2}+{G}\left(\mathbf{r}_{2},\mathbf{r}_{1}\right)\mathbf{p}_{1}\right],\label{Twodip}
\end{eqnarray}
where the Green tensor and the polarizability, in this case, are
\begin{eqnarray}
G({\bf r}_{1},{\bf r}_{2}) & = & {G}({\bf r}_{2},{\bf r}_{1})=\mathcal{G}\mathbb{I}_{2\times2}=\frac{e^{ikr}}{kr}\left(\frac{(kr)^{2}+ikr-1}{(kr)^{2}}\right)\mathbb{I}_{2\times2},\nonumber \\
\mathbf{\alpha_{i}} & = & \left(\begin{array}{cc}
\alpha_{i} & \alpha_{iM}\\
-\alpha_{iM} & \alpha_{i}
\end{array}\right)
\end{eqnarray}
where $r$ is the distance between the two dipoles.

The general solution of that system of Eqs. under the influence
of a plane wave linearly polarized along the x-axis, and amplitude
$E_{0}$ at dipole $1$ is given by:

\begin{eqnarray}
p_{1x}&=&\frac{E_0}{\mathcal{D}}\left[\alpha_1+\mathcal{G}e^{-ikr}(\alpha_2\alpha_1-\alpha_{2M}\alpha_{1M})-\mathcal{G}^2\alpha_2D_1-\mathcal{G}^3e^{-ikr}D_1D_2\right]\nonumber\\
p_{2x}&=&\frac{E_0}{\mathcal{D}}\left[e^{-ikr}\alpha_2+\mathcal{G}(\alpha_2\alpha_1-\alpha_{2M}\alpha_{1M})-\mathcal{G}^2e^{-ikr}\alpha_1D_2-\mathcal{G}^3D_1D_2\right]\nonumber\\
p_{1y}&=&\frac{E_0}{\mathcal{D}}\left[-\alpha_{1M}-\mathcal{G}e^{-ikr}(\alpha_1\alpha_{2M}+\alpha_2\alpha_{1M})-\mathcal{G}^2\alpha_{2M}D_1\right]\nonumber\\
p_{2y}&=&\frac{E_0}{\mathcal{D}}\left[-e^{-ikr}\alpha_{2M}-\mathcal{G}(\alpha_1\alpha_{2M}+\alpha_2\alpha_{1M})-\mathcal{G}^2e^{-ikr}\alpha_{1M}D_2\right],
\label{Dipoles}\end{eqnarray}where $\mathcal{D}=1-2\mathcal{G}^2(\alpha_2\alpha_1-\alpha_{2M}\alpha_{1M})+\mathcal{G}^4D_1D_2$,
and $D_i=\alpha_i^2+\alpha_{iM}^2$. Note that the y-component of
both dipoles is not zero when at least one of the dipoles is MO active.

For the particular geometry we are analyzing, the external magnetic
field produces a change in the polarization state of the reflected
light, and the magneto-optical activity (MOA) of the whole system,
defined as the modulus of the complex Kerr rotation, can be written
as :

\begin{equation}
\text{MOA}=|\theta+i\phi|=\textnormal{atan}\frac{\left|E_{y}^{R}\right|}{\left|E_{x}^{R}\right|}\approx\left|\frac{p_{1,y}+p_{2,y}}{p_{1,x}+p_{2,x}}\right|=\frac{(|p_{1,y}|^{2}+|p_{2,y}|^{2}+2|p_{1,y}||p_{2,y}|\cos(\Delta))^{\frac{1}{2}}}{(|p_{1,x}|^{2}+|p_{2,x}|^{2}+2|p_{1,x}||p_{2,x}|\cos(\Gamma))^{\frac{1}{2}}}.\label{CRot}
\end{equation}

From the interaction point of view there are three different regimes
that are determined by the distance between the interacting dipoles:
strong interaction (very close dipoles), weak interactions (very far
away objects) and medium interaction (intermediate distance). We will
concentrate on the most interesting case of medium interactions,\cite{armelles2013mimicking}
and will analyze two situations: one in which the amount of Co in the
magneto-optical disk is very small (0.1\%) and a second one where it is comparable to the Au
amount (25\%). For the analysis, the aspect ratios of the dipoles
are $a/c=13$ and $10$ for the bottom and top dipoles respectively.

\begin{figure*}
\includegraphics[width=1\textwidth]{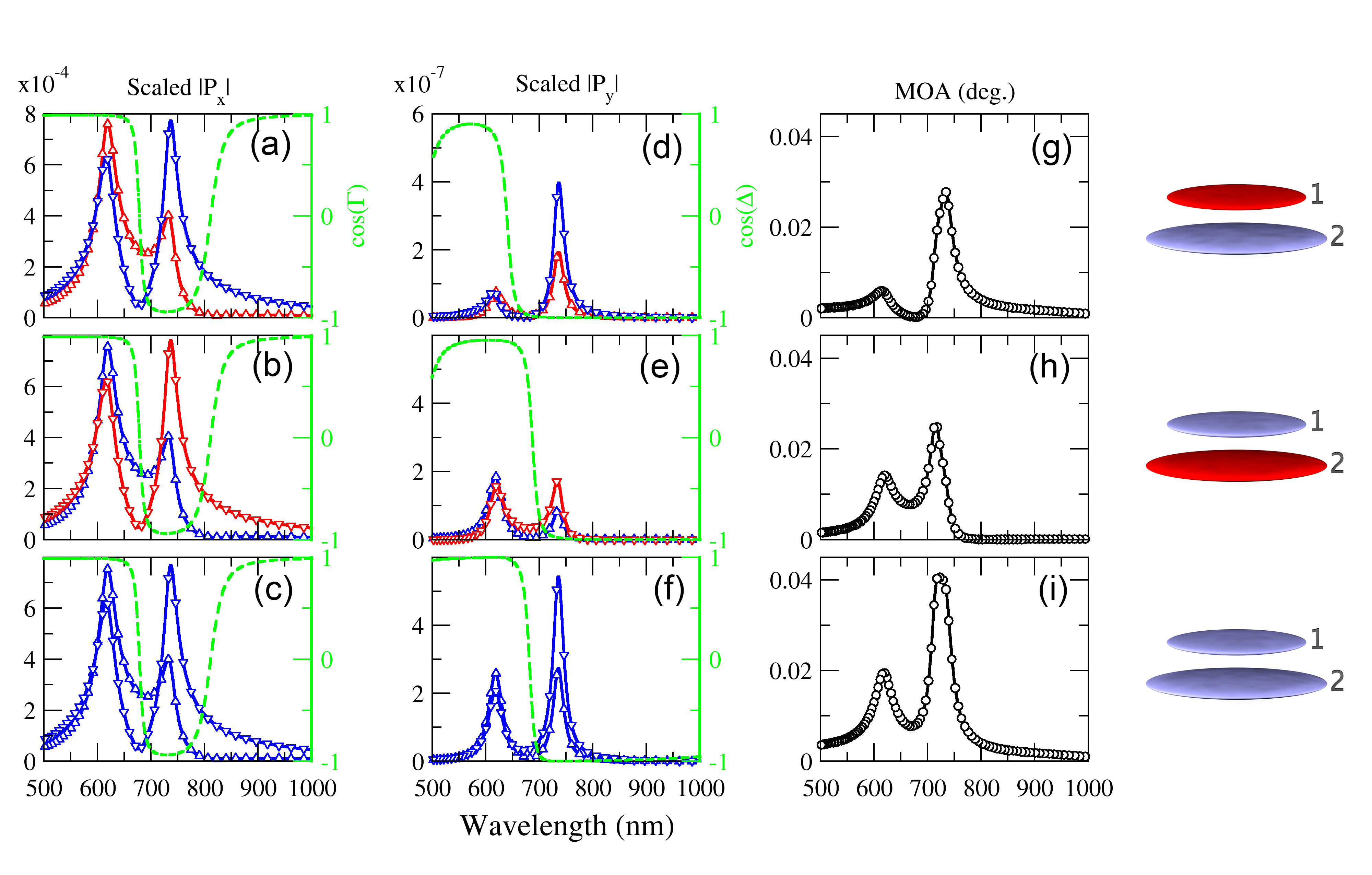}

\caption{Color online. Dipole contributions, and MOA for 0.1\% Co concentration. (a)-(c)
x-component of the scaled dipole (left axis) and the cosine of the
relative phase between them (right axis), (d)-(f) y-component of the
scaled dipole (left axis), and their relative phase (right axis),
(g)-(i) magneto-optical activity. The upper panels represent the situation
where the MP dipole is at the bottom, the medium panels are when the
MP dipole is at the top, the lower panels when both are MP. Triangle-up
for the top dipole, and triangle down for the bottom dipole.\label{LowInt}}
\end{figure*}
Let us start with the case of very small Co concentration (0.1\%)
in the MP dipole. In Fig. \ref{LowInt} we show the modulus of the
components of the dipole along $x$, the polarization direction of
the incident beam ($p_{i,x}$), and along $y$ direction ($p_{i,y}$),
as well as that of the complex Kerr rotation (MO activity, MOA) calculated
with this simple analytical model for the situations where the MP
dipole is at the bottom, top and in both positions of the structure.
The cosine of relative phases between the $p_{i,x}$ ($\cos\Gamma$
in Fig. \ref{LowInt} (a)-(c)) and $p_{i,y}$ ($\cos\Delta$ in Fig.
\ref{LowInt}(d)-(f)) components of the upper and lower disks (dashed
curves) are also shown, with limit values of 1 and -1 for in-phase
and out-of-phase oscillations, respectively.

Considering first the x-component of the dipoles (Fig. \ref{LowInt}
(a)-(c)), and due to the low Co concentration, there is no noticeable
difference between the three situations. All cases show two characteristic
low- (740 nm) and high- (620nm) energy modes of antisymmetric and
symmetric nature respectively, \cite{dmitriev2007gold,armelles2013mimicking,banthi_high_2012}
as directly concluded by the obtained relative phase between the two
dipoles. The abrupt change in sign of the cosine occurs exactly at
the minimum in magnitude of both dipoles. For energies below roughly
760 nm the phase gradually changes again, going back to an in-phase
configuration for wavelengths larger than 820 nm. Regarding the relative
contribution of both disks to these $p_{x}$ spectral features, the
low energy mode has a stronger component originating from the bottom
disk (down-triangles) than from the top one, (up-triangles), since
it has a lower aspect ratio. The situation is reversed for the high
energy peak, even though the difference between the contributions
of the two dipoles is smaller.

Beyond the purely optical properties, fully understandable by simply
considering $p_{x}$, the direct consequence of the application of
a magnetic field is the generation of a y-component in the dipole
\cite{sepulveda2010gold,armelles2013mimicking} (Fig. \ref{LowInt}(d)-(f)).
Contrary to what is observed in the x-component, now different results
are obtained depending on the specific position of the MP active dipole. 

Let us examine each situation individually. When the MP dipole is
at the bottom, a y-component is observed not only in this dipole,
but also in the P top one, which is due to the interaction between
the dipoles. This y-component is stronger for the bottom MP dipole
in the low energy region, but they are similar in the high energy
region. Even more, in the spectral region where the x-component of
the MP dipole is minimum, the y-component of {\em both} dipoles is
almost zero, even though the x-component of the P dipole in the same
intermediate region is not negligible. This is simply due to the fact
that the y-component is originated by the magnetic field induced rotation
of the MP dipole, which in turn induces the rotation of the upper
P dipole. Thus, a y-component dipole can be originated {\em only}
if the x-component of the MP active dipole is not zero. Additionally,
the relative phases between the two dipoles along the y-axes show
essentially the same symmetric/antisymmetric configuration for the
corresponding high/low energy modes compared to those for the x-components,
even though now they do not return to in-phase values for energies
below 800 nm. The presence of $p_y$ is directly related to the presence
of MO activity in the system (Fig. \ref{LowInt} (g)). Indeed, as
shown in Eq. \ref{CRot}, this magnitude is basically the modulus
of the sum of the y-components of the top and bottom dipoles divided
by that of the x-components. Therefore, the spectral dependence of
MOA can be understood in simple terms considering these four dipole
components, taking into account their relative phases. So, in this
first considered case with the bottom MP dipole, the high energy peak
results from the addition of both (top and bottom dipoles) y-components,
while the low energy one results from the corresponding difference,
since in this energetic range the y-components are in phase opposition. 

If we consider now the situation where the MP dipole is on the top
of the structure, the results are very different. Strikingly, here
the obtained y-component in the low energy region is larger for the
P dipole than for the MP one, while both components are similar for
the high energy region. This means that the contribution to the MOA
(Fig. \ref{LowInt} (h)) coming from the P dipole in the low energy
region is actually stronger than that of the MP one. This is simply
due to the larger x-component of the P dipole in the low energy region,
which also explains why the y-components in the high energy region
are of similar magnitude for both MP and P dipole in the previous
case. On the other hand, regarding the intermediate spectral region,
and due to the non vanishing x-component of the MP dipole in this
range, both the y-components (especially of the P dipole) and the
MOA are not zero. 

Finally, if both disks have MO component the intensity of the $p_{y}$
components increases for both dipoles and within each mode they follow
the same trend as the corresponding x-component. Besides, for all
the energy regions the intensity of the MOA is larger than that of
the other two configurations. 

\begin{figure*}
\includegraphics[width=1\textwidth]{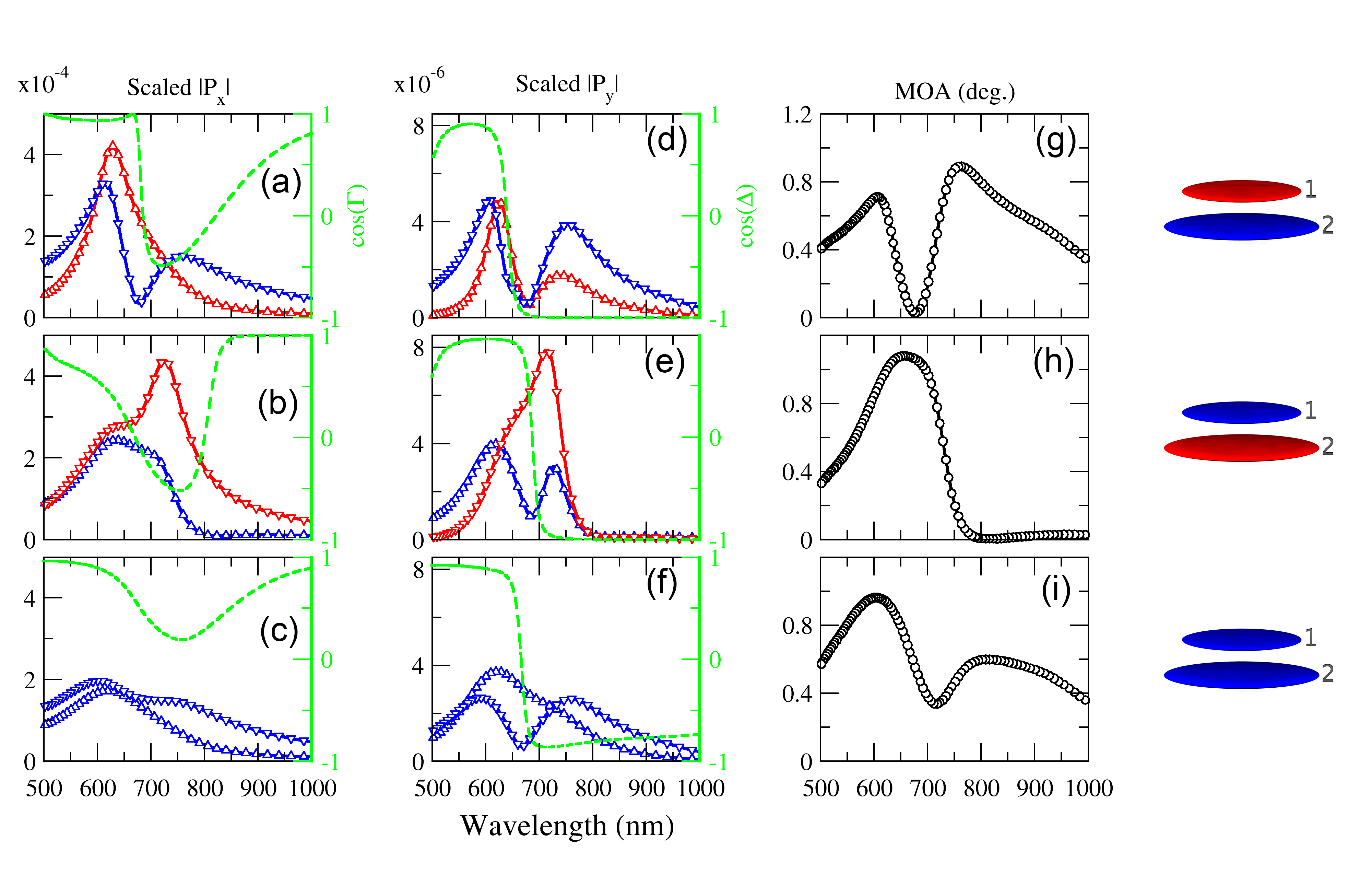}

\caption{Color online. Dipole contributions, and MOA for 25\% Co concentration. (a)-(c) x-component
of the scaled dipole (left axis) and the cosine of the relative phase
between them (right axis), (d)-(f) y-component of the scaled dipole
(left axis), and their relative phase (right axis), (g)-(i) magneto-optical
activity. The upper panels represent the situation where the MP dipole
is at the bottom, the medium panels are when the MP dipole is at the
top, the lower panels when both are MP. Triangle-up for the top dipole,
and triangle down for the bottom dipole \label{HighInt}}
\end{figure*}

Going towards a more realistic situation, with larger Co amounts in
the MP dipoles, in Fig. \ref{HighInt} we show theoretical calculations
equivalent to those shown in Fig. \ref{LowInt} but using the dipole
model with a 25\% Co content in the different MP dipoles. As it can
be seen in Fig. \ref{HighInt}(a)-(c), and contrary to what was
observed for low Co concentrations, now the $p_{x}$ components are
very different depending on the specific position MP dipole. The effect
of increasing the Co amount is both to broaden the peaks and to change
their absolute and relative intensities, as well as their energetic
position, both for $p_{x}$ and $p_{y}$ components. Due to the much
larger amount of Co (250 times more Co) in this case, the magnitudes
of the $p_{x}$ and $p_{y}$ components are now very different (between
a factor of 2 to 4 reduction in the x-component due to the increased
losses, and roughly one order of magnitude larger in the y-component
due to the much larger amount of MO material). Even more, for this
concentration, all these effects also depend on the specific location
of the MP dipole.

For the configuration with the MP dipole at the bottom, the low energy
peak in the MOA has a stronger component due to the bottom dipole
(as seen in the low Co concentration case) and the increase of the
Co amount brings as a consequence both a reduction of the relative
intensity for the x-component and a broadening for both x- and y-components.
However, the high energy peak, with a stronger contribution from the
upper P dipole, is less affected since no Co is present in it. Again,
a minimum in the y-component in the intermediate energy region yields
a minimum in the MO activity. Regarding the relative phases, for the
$p_{x}$ components it is clear now that they do not reach the perfect
out-of phase configuration, indicating that now the nature of the
modes is not purely but only partially antisymmetric. This is due
to the sizeable amount of Co present in the dipoles, which enhances
the losses of the system and affects the retardation between the two
coupled dipole x-components. However, for the y-components, the phase
basically reproduces the same behavior observed for the very low concentration
limit. 

Going now to the situation where the MP dipole is in the upper part,
the most affected peak (for all $p_{x}$, $p_{y}$ and MOA) due to
the incorporation of Co is the high energy one, since it is the one
which carries a stronger part of the upper dipole. We therefore observe
a change in the relative intensity with respect to the case with the
MP dipole in the bottom. Remarkably, in this situation the y-component
of the P dipole is much larger than that of the MP one in the low
energy part of the spectrum, due to the interaction effects and to
the very large x-component of this P dipole. Now, due to both the
broadening and spectral overlapping of the y-components of both top
and bottom dipoles, only one broad peak is observed in the MOA. This
peak is mainly originated by the induced y-component in the bottom
dipole, which is not MO active. Regarding the phase of the y-component,
it is worth noticing that it is again exactly the same as for the
low Co concentration, i.e. it does not depend on the Co concentration.
If one considers Eq. \ref{Dipoles}, and makes either $\alpha_{1M}$
or $\alpha_{2M}$ equal to zero, the ratio between the y-components
of the P and MP dipoles becomes $p_{y,NMO}/p_{y,MO}=\mathcal{G}\alpha_{NMO}$,
i.e. it does not depend on the MO active element, and thus the relative
phase does not depend on the Co content. 

Finally, when both dipoles are MP, the larger amount of Co implies
that the total losses are even larger, and therefore the peaks in
the x-components are weaker and broader. The direct consequence of
this reduction in the x-components is that the y-components are also
somehow weaker and broader compared to the other two cases with only
one MP dipole. Now, for the x-component the two peaks overlap for
the bottom dipole and only one broad peak is observed for the top
one, which is also the same behavior basically observed for the y-component.
Regarding the MOA, two well distinguishable peaks are observed and,
surprisingly, the low energy one yields lower MO activity than the
corresponding peak for the other two cases (only one MP dipole), contrary
to what was observed for low Co concentration. This is due to the
smaller difference of intensities between the y-components of the
two dipoles in this case and to the fact that in this spectral region
they are out-of-phase. Briefly, this complex behavior of relative
intensities and phases induces a smaller MOA even when more MP components
are present in this dimer.

Let us summarize the preceding discussion. When two dipoles interact,
one of them presenting MO activity (MP dipole), this MP dipole can
induce MOA in the non-MO active (purely plasmonic, P) one. The induced
MOA can be even much larger than the intrinsic one (see low energy
region of Figs. \ref{LowInt}e and \ref{HighInt}e). This occurs
in the spectral region of the resonance of the P dipole. On the other
hand, if the system is composed of two MP, lossy dipoles (high Co
content), the resulting MOA response can be much lower than that obtained
when only one dipole is MP (see Figs. \ref{HighInt}g-i). This has
important consequences in real magnetoplasmonic systems composed of
noble metals and ferromagnetic metals. One would naively think that
the MOA is to be enhanced by increasing the number of MP components,
but then the losses will increase in parallel. Our results show that
an adequate stacking of the system components may allow devising structures
with higher MO activity using overall lower amounts of ferromagnetic
content.

%\section{Experimental results and numerical simulations for real stuctures}

\begin{figure}
\includegraphics[width=1\columnwidth]{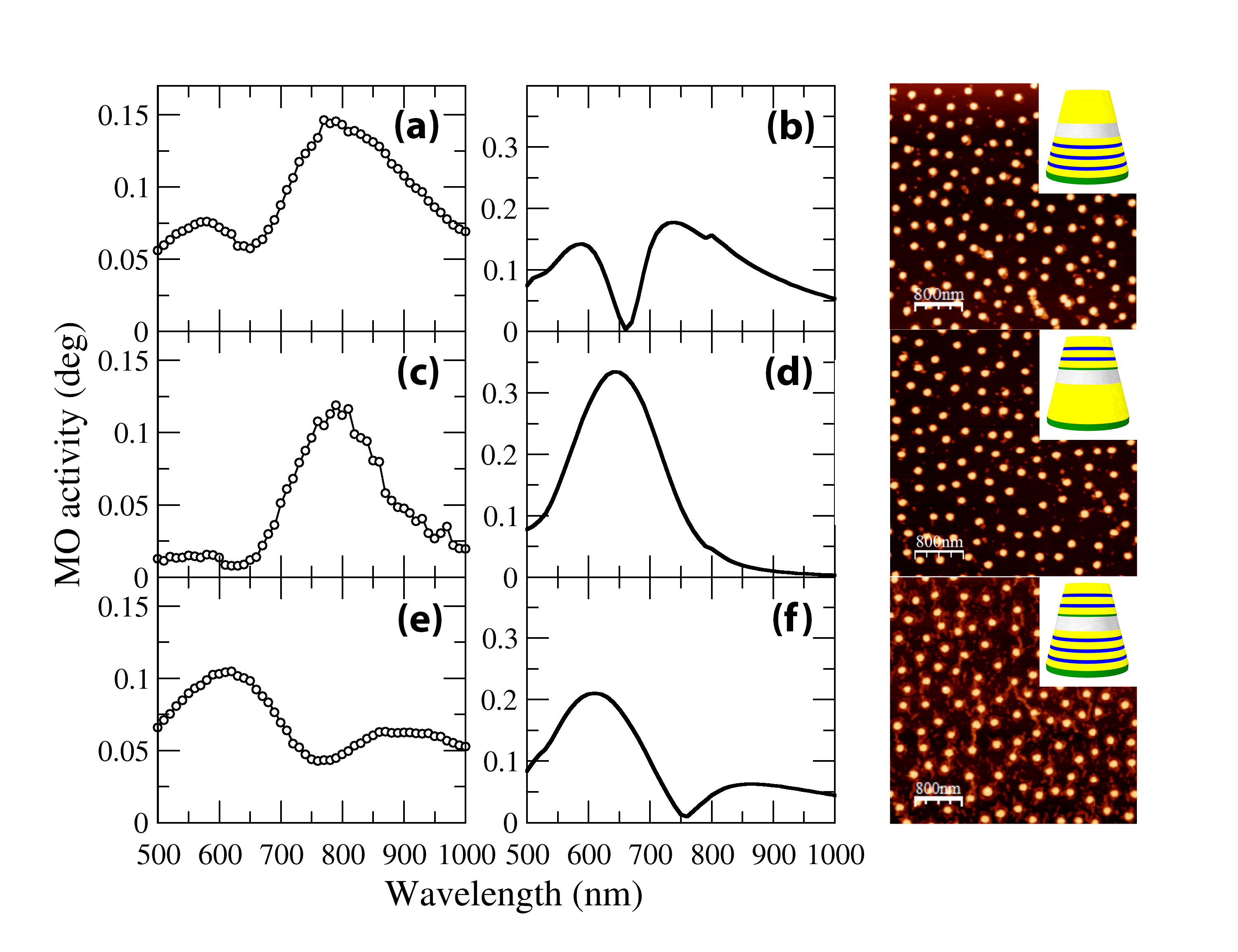}

\caption{Color online. Experimental results (a) and numerical simulations (b) of the MO activity
when the MP disk is at the bottom, (c),(d) are the same but when the
MP disk is at the top, and (e),(f) when both are MP. In the rightmost panel we show AFM images of the three experimental samples where the density of disks (about 15\% coverage) and homogeneity can be seen. The images show that the disk diameter ranges from 130nm to 150nm. Also is represented a scheme of the structures. In green we
depict the Ti adhesive layer, in blue the Co layers, grey is the SiO$_{2}$
and yellow are the gold layers. The three different systems are constituted
by two metallic disks separated by 20nm of SiO$_{2}$. For the configuration
with the MP disk at the bottom (top panels), the MP disk is composed
of a 2nm Ti layer followed by a 4nm Au layer and three sequential
combinations of 2nm Co/4nm Au layers. The disk at the top is composed
by 16nm Au. For the configuration with MP disk at the top, the MP
disk is composed by an initial 1nm layer of Ti then 4nm Au layer and
two sequences of 2nm Co/4nm Au layers. The disk at the bottom is composed
by 22nm of gold. When both disks are MP, they consist of those Au/Co sequences employed in the other two situations. \label{Experim}}
\end{figure}
Despite the simplicity of the two interacting dipoles model, it describes
quite well the outcome of the interaction between disks in magnetoplasmonic
dimers. For example, in Fig. \ref{Experim} we present the experimental
MO activity for three different samples. They consist of a layer of
two metallic disks separated by 20nm of SiO\textsubscript{2} deposited
on a glass substrate. The three samples have a homogeneous distribution of the disks, with a filling factor of 15\%. 
The diameter of the disks ranges from 130nm to 150nm. They were obtained by hole mask colloidal lithography,
metal evaporation and lift off. \cite{dmitrievcolloidal} The internal
structure of the disks is presented in the rightmost panel of Fig.
\ref{Experim}. The disks dimensions are the same as those of the
disks in Ref. \cite{armelles2013mimicking}. In sample a, the bottom
disk consists of a Au/Co multilayer (MP) and the top one is a Au disk
(P); in sample b, the top disk is a Au/Co multilayer (MP) and the
bottom one a Au disk (P); and finally, in sample c, both disks are
Au/Co multilayers (MP).

As it can be observed, when the bottom disks are magnetoplasmonic,
samples a and c, the MOA spectrum has two peaks. Despite the lower
Co contents of sample a, the lower energy peak has a higher intensity
in this sample than in sample c, where the two disks are magnetoplasmonic.
Moreover, the MOA spectrum of sample b has only one peak, whose intensity
is also greater than the intensity of the low energy peak of sample
c. Additionally, in Fig. \ref{Experim} we also present a FDTD theoretical
calculation which takes into account the internal structure of the
disks. As it can be observed, these calculations reproduce quite well
the experimental behavior, and the results are also equivalent to
those obtained with the previously exposed analytical approach for
intermediate interactions (Fig. \ref{HighInt}). The numerical calculations have 
been made using 130nm for the diameter of base of the cone and 100nm for the top in all cases.
An increase in these numbers would cause a red shift of the peaks.

%\textbf{Conclusions and outlook}

In conclusion, we have analyzed the effect of electromagnetic interactions
on the MO response of magnetoplasmonic dimers composed of two metallic
disks separated by a dielectric. The MO response strongly depends
on the plasmonic versus magnetoplasmonic nature of the two disks,
observing for specific configurations that the MO response can be
dominated by the induced MOA of the purely plasmonic disk. On the
other hand, the MO activity of a system with only one of the disks
containing material with intrinsic MO can be even larger than that
of a system composed of two MP disks. A simple analytical model of
two interacting point dipoles allows us to fully describe separately
the contribution of each disk to the optical and MOA of the system,
along with the relative phases of the dipoles responsible for these
activities. Experimental results and numerical calculations fully
support the analytical calculations results. 

\begin{acknowledgments}

We acknowledge H. Feng for his assistance in the growth of some of
the studied samples. Funding from Spanish Ministry of Economy and
Competitiveness through grants \textquotedblleft{}FUNCOAT\textquotedblright{}
CONSOLIDER CSD2008\textendash{} 00023, and MAPS MAT2011\textendash{}29194-
C02\textendash{}01, and from Comunidad de Madrid through grants \textquotedblleft{}NANOBIOMAGNET\textquotedblright{}
S2009/MAT-1726 and \textquotedblleft{}MICROSERES-CM\textquotedblright{}
S2009/TIC-1476 is acknowledged. L.S. F.-P. acknowledges support from
the European Social Fund and CSIC through a JAE-Doc grant.
\end{acknowledgments}

\bibliography{bibliography,Bibliopolitica}
\end{document}